\documentclass[12pt]{iopart}
\usepackage{iopams} 
\usepackage{graphicx} 
\newcommand{\ybco}{$\mathrm{YBa_2Cu_3O_{7-\delta}}$}
\newcommand{\ybcoo}{$\mathrm{YBa_2Cu_3O_{6.93}}$}
\newcommand{\ybcooo}{$\mathrm{YBa_2Cu_3O_{7}}$}
\newcommand{\bscco}{$\mathrm{Bi_2Sr_2CaCu_2O_8}$}
\newcommand{\rr}{{\bf r}}
\newcommand{\kk}{{\bf k}}
\newcommand{\KK}{{\bf K}}

\begin{document}

\title[Effect of Chains on the LDOS in the Vortex Phase of YBCO]{Effect of CuO
chains on the local density of states in the vortex phase of
$\mathrm{YBa_2Cu_3O_7}$}

\author{W A Atkinson} 

\address{Department of Physics and Astronomy, Trent University, 1600
West Bank Dr., Peterborough ON, K9J 7B8, Canada}

\ead{billatkinson@trentu.ca}

\begin{abstract}
We examine the effects of the CuO chains on the density of states in
the vortex phase in $\mathrm{YBa_2Cu_3O_7}$, via a calculation based
on the tight-binding proximity model.  In this model, chain
superconductivity results from single-electron hopping between the
intrinsically-normal chains and intrinsically-superconducting CuO$_2$
planes.  The calculations are based on self-consistent solutions of
the Bogolyubov-de Gennes equations for a bilayer consisting of a
single CuO$_2$ layer and a single CuO chain layer.  We find that, in
addition to the dispersing resonances found in single-layer models,
the chains introduce a second set of dispersing resonances associated
with the induced gap in the chain layer.  These new resonances are
highly anisotropic and distort the vortex core shape.
\end{abstract}

\submitto{\SUST}


\maketitle

\section{Introduction}
Because of the availability of large high-quality single crystals,
\ybco{} is one of the most widely studied of the high temperature
superconductors (HTS).  As with other HTS, the essential structural
components are the two-dimensional (2D) CuO$_2$ layers.  In these
layers, the Cu and O orbitals are partially filled and are therefore
conducting.  The interlayer coupling is generally weak, meaning
that the electronic bands contributing at the Fermi energy derive most of
their weight from individual CuO$_2$ layers.  For this reason, the
standard model for a generic cuprate HTS consists of a single 2D
CuO$_2$ layer in isolation.

\ybco{} differs from other cuprate HTS in one significant respect: in
addition to the CuO$_2$ layers, there are layers of one-dimensional
(1D) chains.  Band structure calculations\cite{Andersen1995} suggest
that the chains are roughly one-quarter filled, and contribute a
quasi-1D Fermi surface.  Despite this, the chains are nearly always
ignored in theories of \ybco, presumably because it has many
physical properties that are common to all HTS, suggesting a minor
role for the chains.  In addition, the metallicity of the chains has
been questioned, either because electrons in 1D are localized
by arbitrarily weak disorder, or because they are unstable towards the
formation of insulating charge density wave states.  

Early evidence for the metallicity of the chains came from
measurements of the resistivity anisotropy\cite{Friedmann1990}, which
found $\rho_a/\rho_b \approx 2$ at 100 K, where $\rho_a$ and $\rho_b$
are the resistivities parallel to the two axes of the CuO$_2$ planes,
and where $b$ is also parallel to the chains.  Such an anisotropy is
expected if the chains are conducting and carry current in parallel
with the planes.  Since O-vacancies, which are prevalent in \ybco,
would strongly localize 1D electrons, the observed anisotropy appears
to indicate a relatively large hybridization of chain and plane
states.  Further support for this comes from c-axis (perpendicular to
the planes) resistivity measurements\cite{Hagen1988,Takenaka1994}
which find an anisotropy $\rho_c/\rho_a\approx 50$ at 100 K in
optimally-doped \ybcoo.  In a tight binding model, this corresponds to
$t_\perp/t \approx 7$ where $t$ and $t_\perp$ are the in-plane and
perpendicular hopping amplitudes.  This anisotropy is smaller than in
other HTS and suggests coherent $c$-axis transport\cite{Homes2005}.

There has been some debate as to whether the CuO chains undergo a
Peierls transition to a charge density wave state.  Scanning tunneling
microscopy (STM) experiments on chain-terminated surfaces found charge
modulations which were interpreted as charge density waves.\cite{Edwards1995}
However, later experiments\cite{Derro2002} found that the modulation
wavelength depends on the bias voltage of the STM tip, which is a
characteristic of Friedel oscillations (i.e.\ standing wave patters
produced by impurity scattering of itinerant electrons) rather than
charge density waves.  These later experiments are consistent with
metallic chains.

Direct evidence for a chain Fermi surface has recently been found in a
number of angle resolved photoemission
experiments\cite{Lu2001,Zabolotnyy2007,Nakayama2007}.  The Fermi
surface appears to be consistent with that predicted by band structure
calculations, although a complete characterization of the band
structure is rendered difficult by the existence of surface states,
which dominate surface-sensitive experiments such as tunneling and
photoemission.  The correspondence between the surface states and
electronic states in the bulk is not yet established.

Complementary information on the chain states comes from penetration
depth anisotropy measurements\cite{Zhang1994,Basov1995} which indicate
that the chains are superconducting below the bulk critical
temperature $T_c$, and have a substantial superfluid density.  Because
the chains and CuO$_2$ planes are structurally different, the apparent
similarity of their superconducting states, as measured by the
penetration depth, was puzzling for many years.  The chains, being far
from half-filled, should not have an intrinsic pairing interaction
capable of producing high critical temperatures.  A more likely
scenario is that the chains derive their superconductivity from
proximity coupling to the CuO$_2$ planes.\cite{Atkinson1995a}
Calculations showed that proximity models for chain superconductivity
introduce small energy scales related to the induced gap in the
chains.  To date, these small scales have not been observed in
penetration depth measurements.\cite{Atkinson1995b} It was later
shown\cite{Atkinson1999} that the smallest energy scale, which is
relevant to low temperature measurements, comes from a subset of
chain-derived electronic states that are weakly hybridized with the
CuO$_2$ planes.  Since these particular states have a predominantly 1D
character, they are are strongly affected by localization corrections
due to chain disorder, and are therefore difficult to detect via the
electrodynamic response.

Small energy scales have been seen in other experiments.  For example,
it has been found\cite{Sonier1999b,Sonier2004} that the vortex cores
in \ybco, as measured by $\mu$SR experiments, are much larger at small
magnetic fields $B$ than expected from the measured upper critical
field, and that the cores contract rapidly with increasing $B$ below a
crossover field $B^\ast$.  This anomalous behaviour has been explained
by the presence of CuO chains\cite{Sonier2007,Atkinson2008}, and it
can be shown that $B^\ast \sim E_\mathrm{s}^2$.  It has also been
suggested that a similar small energy scale seen in tunneling
experiments\cite{Valles1991,Yeh2001,Ngai2007} can be attributed to the
chains.

In summary, we argue that there is reasonable evidence that the CuO
chains in \ybco{} (a) are metallic, (b) are signifcantly hybridized with
the CuO$_2$ planes, and (c) superconduct as a result of this
hybridization.  There remain many unresolved questions regarding the
role of the chains, however.  First, there are practical issues:
neither the strength nor momentum-dependence of the chain-plane
coupling have been established with any degree of certainty.  Second,
there are questions of fundamental interest: what are the properties
of a weakly correlated 1D metal in close contact with a
strongly-correlated electron liquid.  Finally, there is the question
of how the chains manifest themselves in various experiments.  In some
cases, for example transport anisotropy, the role played by the chains
is intuitively obvious.  However, in many cases, the effects of the
chains are not {\em a priori} obvious, as in the case of the vortex
core contraction mentioned above.

\begin{figure}
\begin{center}
\includegraphics[width=0.6\columnwidth]{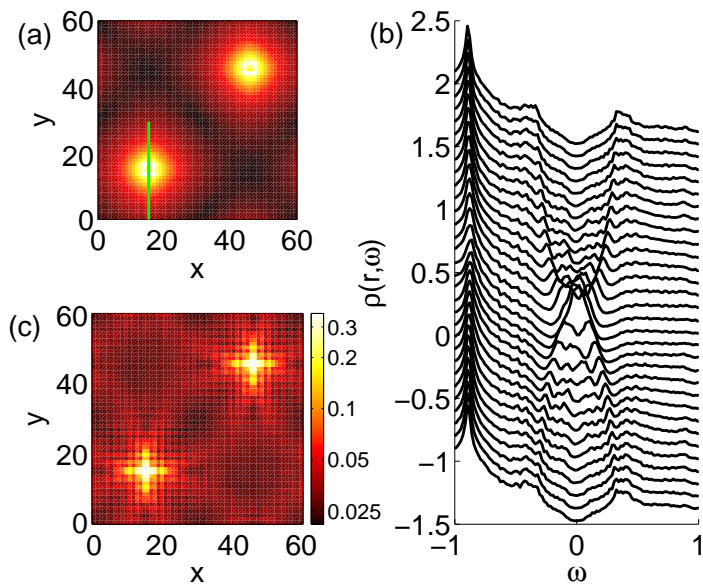}
\caption{Electronic structure near the vortex core for a single
tetragonal layer with a $d$-wave superconducting order parameter.  (a)
Magnitude of the current density $|{\bf j}({\bf r})|$ as a function of
position.  The figure shows a pair of vortices belonging to a square
lattice whose axes are rotated $45^{\circ}$ relative to the
crystalline axes (which are parallel to the figure axes).  (b) Local
density of states at a series of positions along the line drawn
through the lower vortex core in (a).  Note that the nondispersing
peak at $\omega = -0.9$ is a van Hove singularity associated with the
band structure.  Far from the vortex core, the $d$-wave order
parameter is $|\Delta| = 0.38$.  (c) The LDOS
$\rho(\rr,\varepsilon_F)$ is shown as a function of position near a
single vortex core. Note the logarithmic scale.  Here and throughout
the Fermi energy is $\varepsilon_F=0$.}
\label{fig:cut1}
\end{center}
\end{figure}

In this work, we examine the effects of chains on the local density of
states (LDOS) $\rho(\rr,\omega)$ in the vortex state of \ybcooo.
Experiments have found that the electronic structure of vortices in
\ybco{} is different from other HTS\cite{Fischer2007} and from
theoretical calculations\cite{Wang1995,Schopohl1995}.  Theory predicts
that, for a single superconducting layer with a $d$-wave order
parameter, there should be a peak in the LDOS at the vortex core
centre (a ``zero bias conduction peak'' or ZBCP), and that as one
moves away from the core, the peak splits and disperses towards higher
energies, eventually merging with the coherence peaks far from the
vortex core.  This behaviour is illustrated in Figure \ref{fig:cut1},
where the results of a calculation performed for a single-layer
version of the model used in this work are shown.

Experimentally, the situation is rather different. The ZBCP at the
vortex core is absent, and a pair of finite-energy peaks at $\omega
\approx \pm 5$ meV are present instead.  Furthermore, unlike the
calculations, these peaks do not shift significantly as one moves away
from the vortex core.  It is possible that these peaks have the same
origin as peaks measured in zero-field tunneling experiments, but
this is not established.

In this work we ask two questions: (i) what are the signatures of the
chains in the vortex structure within the proximity model and (ii) to
what extent can the proximity model explain the existing experiments?

\section{Method}

Since the technical details of the calculation have been given in
\cite{Atkinson2008}, we only give a summary of the main ideas here.
The plane-chain model is made up of a single bilayer consisting of a
tetragonal 2D plane layer (representing a CuO$_2$ layer) and a layer
of 1D chains (representing a CuO layer).  The layers are connected by
single electron hopping, so that the plane and chain bands hybridize
to form new bands. These bands have either predominantly plane or
chain character, and we refer to them as plane-derived or
chain-derived respectively.  The chain layer is represented by a
tight binding Hamiltonian with a single orbital per unit cell. In band
structure calculations, this orbital is a linear combination of the
chain Cu$_{d_{z^2-y^2}}$ and O$_{p_{y}}$ orbitals.\cite{Andersen1995} The
CuO$_2$ band structure is also represented by a tight binding model
with a single orbital per unit cell.  In this case, the orbital
represents a singlet state involving the Cu$_{d_{z^2-y^2}}$ and
$O_{p_{(x,y)}}$ orbitals.\cite{Zhang1988}.

There is a pairing interaction in the plane layer leading to $d$-wave
superconductivity, but the chain layer is intrinsically normal.
Nonetheless, the chain superconducts because of the the proximity
effect; single electron hopping between the plane and chain layers
induces a gap in the chains.  One important feature of the proximity
model is that the induced gap in the chains does not share the
$d$-wave symmetry of the plane layer\cite{Atkinson1999}.  This is
because the chain dispersion is quasi-1D, and the $\kk$-dependence
of the excitation gap depends on the symmetry of both the order parameter 
and the underlying band.

We define coordinates $x$, $y$, and $z$ which are aligned with the
crystalline axes $a$, $b$, and $c$ respectively, and consider a
magnetic field aligned with the $c$-axis.  The screening currents,
therefore, circulate within the plane and chain layers.  It is for
this configuration that the vortex core contraction is seen in $\mu$SR
experiments,\cite{Sonier1999b} and that STM experiments are performed.

The  Hamiltonian is
\begin{equation}
\hat H = \hat H_1 + \hat H_2 + \hat H_\perp.
\end{equation}
where $\hat H_1$ is the Hamiltonian for the isolated plane, 
$\hat H_2$ the Hamiltonian for the isolated chains, and 
$\hat H_\perp$ the single-electron hopping term that couples the two layers.
For comparison, we also consider a single-layer model described by
$\hat H_1$ alone.
We have
\begin{eqnarray}
\hat H_1 = \sum_{ij\sigma} \tilde t_{1ij}
c^\dagger_{1\sigma}(\rr_i) c_{1\sigma}(\rr_j) 
+ \sum_{ij} [
\Delta_{ij} c^\dagger_{1\uparrow}(\rr_i)
c^\dagger_{1\downarrow}(\rr_j) 
\nonumber \\  
\qquad +\Delta_{ij}^\ast
c_{1\downarrow}(\rr_j) c_{1\uparrow}(\rr_i) ],
\end{eqnarray}
where $c_{1\sigma}(\rr_i)$ is the annihilation operator for an
electron in the plane on site $i$ with spin $\sigma$, and position
$\rr_i=(x_i,y_i)$, $\tilde t_{1ij}$ are hopping matrix elements, and
$\Delta_{ij}$ are superconducting pair energies.  The subscripts ``1''
and ``2'' refer to the plane and chain layers
respectively.  The hopping matrix element $\tilde t_{1ij}$ between
sites $i$ and $j$ includes the effects of the magnetic field via
\begin{eqnarray}
\tilde t_{1ij} =
t_{1ij}\exp\left [-\rmi\frac{e}{\hbar c}\int_{\rr_j}^{\rr_i} \rmd\rr\cdot
{\bf A}(\rr) \right] \nonumber \\
= t_{1ij} \exp \left[ \rmi\alpha \frac{y_i+y_j}{2} (x_i-x_j) \right ],
\label{Eq:tij}
\end{eqnarray}
where $t_{1ij}$ are the zero-field matrix elements, ${\bf
A}(\rr_i) = -B_0y_i{\bf \hat x}$ is the static magnetic vector
potential,  $B_0$ is the uniform applied magnetic field, and
$\alpha = eB_0/\hbar c$.

We take a square tight-binding lattice with hopping up to
second-nearest neigbours.  The matrix elements in zero magnetic field
are $t_{1ii} = t_{1,0}$, $t_{1\langle i,j\rangle}=t_{1,\mathrm{nn}}$,
and $t_{1\langle \langle i,j \rangle \rangle} = t_{1,\mathrm{nnn}}$,
where $\langle i,j\rangle$ and $\langle\langle i,j\rangle \rangle$
refer to nearest and next-nearest neighbours respectively.  When
$B=0$, the dispersion is $\epsilon_1(\kk) = t_{1,0} +
2t_{1,\mathrm{nn}}( \cos k_x + \cos k_y) + 4t_{1,\mathrm{nnn}}\cos k_x
\cos k_y$.

The order parameter for superconductivity is denoted $\Delta_{ij}$,
where $i$ and $j$ denote sites on the lattice.  The order parameter is
determined self-consistently using a nearest-neighbour attractive
interaction of magnitude $V$.  Then,
\begin{equation}
 \Delta_{ ij } = -\frac {V}{2} \langle
c_{1\downarrow}(\rr_j) c_{1\uparrow}(\rr_i) + c_{1\downarrow}(\rr_i)
c_{1\uparrow}(\rr_j) \rangle \delta_{\langle i,j \rangle}.
\end{equation}
 The d-wave component, defined by
$\Delta(\rr_i) = \sum_j (-1)^{y_i-y_j}\Delta_{ij},$
is the dominant component of the order parameter.

The Hamiltonian for the chain layer is 
\begin{equation}
H_2 = \sum_{ij\sigma} t_{2ij} c^\dagger_{2\sigma}(\rr_i)
c_{2\sigma}(\rr_j)
\end{equation}
where $t_{2ii}=t_{2,0}$ and $t_{2ij} = t_{2,\mathrm{nn}}$ for $i$ and
$j$ nearest-neighbour sites belonging to the same chain.  The matrix
elements are unchanged by the magnetic field because of the choice of
gauge.  When $B=0$, the chain dispersion is $\epsilon_2(\kk) = t_{2,0}
+ 2t_{2,\mathrm{nn}}\cos k_y$.  

The term describing the interlayer hopping is 
\begin{equation}
H_\perp = t_\perp \sum_{i\sigma}
[c_{1\sigma}^\dagger(\rr_i)c_{2\sigma}(\rr_i) + c_{2\sigma}^\dagger(\rr_i)
c_{1\sigma}(\rr_i)],
\end{equation}
 which mixes the chain and plane wavefunctions.  

The model parameters used in this work are $\{ t_{1,0},
t_{1,\mathrm{nn}}, t_{1,\mathrm{nnn}}, t_{2,0}, t_{2,\mathrm{nn}},
t_\perp \} = \{ 1.0,-1.0,0.45,2.4,-2.0,0.6 \}$, and the pairing
interaction is $V = 1.3$.  With these definitions, we have taken the
magnitude of the nearest-neighbour hopping
(i.e. $|t_{1,\mathrm{nn}}|$) as the scale of energy.  This will be our
energy scale throughout this work, and a rough comparison to
experiments may be made by taking $t_{1,\mathrm{nn}} \sim 100$ meV.
However, we emphasize that quantitative comparisons to experiments are
not possible because the model parameters are chosen for numerical
convenience (i.e.\ such that the energy scale $E_\mathrm{s}$
associated with chain superconductivity is easily resolved), rather
than to reproduce the \ybco{} band structure accurately.

There is a
quasi-periodicity which allows us to define an $L_x\times L_y$
magnetic supercell containing $N = L_xL_y/a_0^2$ atomic lattice sites
($a_0$ is the lattice constant) and enclosing two superconducting flux
quanta (i.e.\ two vortices), where the superconducting flux quantum is
$\Phi_0 \equiv hc/2e$.  The magnetic field is therefore $B =
2\Phi_0/L_xL_y$.  In order to obtain low magnetic fields, we need
large values of $L_x$ and $L_y$.  Unless otherwise stated, all
results shown in this paper are for a
 $60\times 60\times N_z$-site supercell ($N_z=1,2$) corresponding
to $B = 2\Phi_0/3600$.  For this case, we sum over $5^2 = 25$
$\KK$-vectors, which is formally equivalent to studying a system of 50
vortices on a $300\times 300\times 2$ site lattice.
Taking a unit cell size of $a_0 = 4$ \AA, this
corresponds to $B=7$ T.  While this is a strong field for \ybco{}, we
have chosen the model parameters such that this is in the low-field limit
of the proximity model (i.e.\ chain superconductivity is not quenched
at this field).

The details of the transformation to Bloch
states are given in \cite{Atkinson2008}.  Here, we simply mention that
our self-consistent calculations proceed in several steps.  First, we
diagonalize the Hamiltonian to generate a  set of eigenenergies
$E_{i,\KK}$ and wavefunctions $\Psi_{\alpha,\KK}(n,\sigma,\rr)$ where
$\KK$ are the supercell wavevectors, $\alpha \in (1,2N)$ (the factor of two
is for spin) is a  quantum number labelling the eigenenergies
for each $\KK$, $n = 1,2$ refers to the layer, and $\rr$ to sites
within each layer.  Note that a Bogolyubov transformation has been
made, such that the spin index refers to spin-up electrons with
wavevector $\KK$ ($\sigma=\uparrow$) or spin-down holes with
wavevector $-\KK$ ($\sigma=\downarrow$).  The second step is to
calculate the Fourier-transformed order parameter $\Delta_{ij}(\KK)$
from the wavefunctions.  This new order parameter is inserted into the
Hamiltonian, and a new set of eigenfunctions and energies is
calculated.  The iterative process terminates when the largest
difference between the input and output values of $\Delta_{ij}(\KK)$
is less than $10^{-3}$.  

We present results for two observables in this work.  The LDOS at
energy $\omega$ in layer $n$ is
\begin{eqnarray}
\rho_n(\rr,\omega) = \frac{1}{N_k}\sum_\KK \sum_{\alpha=1}^{2N} 
\left [|\Psi_{\alpha,\KK}(n,\uparrow,\rr)|^2\delta(\omega-E_{\alpha,\KK}) 
\right.
\nonumber \\
\left .
+|\Psi_{\alpha,\KK}(n,\downarrow,\rr)|^2\delta(\omega+E_{\alpha,\KK})\right],
\end{eqnarray}
and the total density of states (DOS) is $\rho(\omega) = N^{-1}\sum_{n}
\sum_\rr \rho_n(\rr,\omega)$ where $N$ is the total number of sites
in the lattice.  The second observable is the 2D current
density in layer $n$, denoted by ${\bf j}_n(\rr) =
(j_{n,x}(\rr),j_{n,y}(\rr))$ with
\begin{eqnarray}
j_{n,x}(\rr_i) = \frac{-e}{2\hbar a_0} \mbox{Im} \frac{1}{N_k} \sum_\KK
\sum_{\alpha=1}^{2N} f(E_{\alpha,\KK}) \nonumber \\
\times \frac 12 \sum_\pm \left [
\tilde t_{n,ij}(\KK) \Psi^\ast_{\alpha,\KK}(n,\uparrow,\rr_i) 
\Psi_{\alpha,\KK} (n,\uparrow,\rr_i\pm a_0{\bf \hat x}) \right.
\nonumber \\
\left .
-
\tilde t_{ij}(-\KK) \Psi_{\alpha,\KK}(n,\downarrow,\rr_i) 
\Psi^\ast_{\alpha,\KK} (n,\downarrow,\rr_i\pm a_0{\bf \hat x}) \right ],
\end{eqnarray}
and a similar expression for $j_{yn}(\rr_i)$.  Note that the layer index $n$
is omitted when we present quantities for the single-layer model.

\section{Results and Discussion}
The LDOS is shown in Figure \ref{fig:cut1} for a model consisting of a
single tetragonal plane.  This case is well-studied, and we include it
here as a point of comparison for later calculations.  In this figure,
one sees a ZBCP at the vortex core.  Away from the vortex core, the
ZBCP splits into two peaks which disperse away from the Fermi energy
$\varepsilon_F$ as the distance to the core increases, eventually
merging with the coherence peaks at $\omega = \pm E_L$.  This is the
behaviour predicted for isolated vortices in both $s$-\cite{Gygi1991}
and $d$-wave\cite{Schopohl1995,Wang1995,Kato2001} superconductors.
For an isolated vortex in an $s$-wave superconductor, quasiparticle
eigenstates have a well-defined angular momentum, and it was shown that
peaks at different distances from the core correspond to different
angular momentum eigenstates.  Sufficiently near the core, both the
peak energy and distance to the core are linearly proportional to the
angular momentum.\cite{Caroli1964} As we will show, this
relationship between angular momentum and the peak energy 
is relevant to the electronic structure of CuO chains in
the vortex state.

We remark that, although dispersing peaks have been observed
experimentally in the $s$-wave superconductor NbSe$_2$\cite{Hess1990},
they have not been seen in either \ybco{} or \bscco.\cite{Fischer2007} A
number of strong-correlation-related mechanisms have been proposed for
the absence of the ZBCP\cite{Fischer2007}, and it is known that
disorder can suppress the ZBCP\cite{Fischer2007}.  Furthermore,
it is possible that nontrivial tunneling matrix elements can significantly
alter the LDOS.\cite{Wu2000}  At present, the
reason for the lack of a ZBCP in the HTS has not been resolved.  

The spatial dependence of the LDOS at the Fermi energy,
$\rho(\rr,\varepsilon_F)$, is plotted in Figure \ref{fig:cut1}(c).  A
logarithmic plot is used to emphasize the long-range tails extending
from the vortex core.  These are a signature of BCS-like $d$-wave
superconductivity\cite{Schopohl1995} and, like the ZBCP, have not been
seen experimentally.  Again, the reason for this is not clear.

\begin{figure}
\begin{center}
\includegraphics[width=0.6\columnwidth]{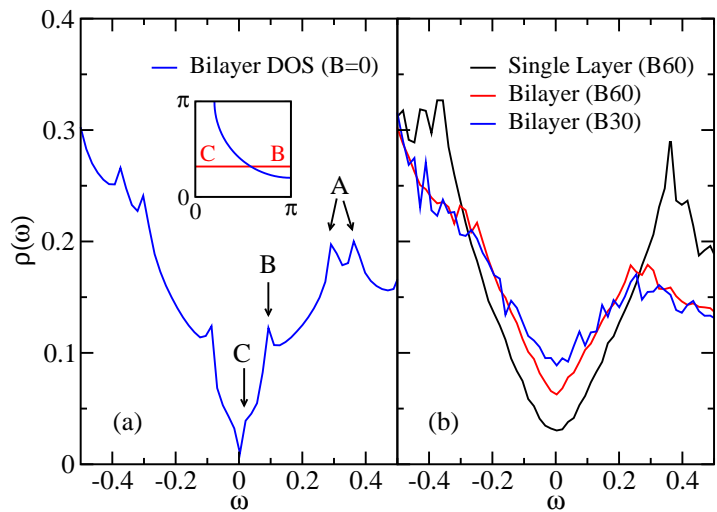}
\caption{Superconducting density of states. (a) Spatially-averaged
density of states in zero magnetic field for the bilayer model.  The
superconducting coherence peaks for the 2D plane are split by
plane-chain coupling (A), while there are two distinct energy scales
(B and C) associated with the induced gap in the chains.  The inset
shows a sketch of the chain (red) and plane (blue) Fermi surfaces.
The portions of the chain Fermi surface contributing to the DOS at the
energies B and C are indicated.  (b) The effect of magnetic field on
the spatially-averaged DOS. The field strengths are $B = 2\Phi_0/60^2$
(B60) and $B=2\Phi_0/30^2$ (B30).  The latter results are calculated
using a $30\times30\times2$ lattice with $15^2$ supercell $\KK$-points.}
\label{fig:cmpr}
\end{center}
\end{figure}

In Figure \ref{fig:cmpr}, we compare the spatially-averaged density of
states  for three different cases: the bilayer model in zero
magnetic field, and the single-layer and bilayer models for $B\neq 0$.
This figure shows that the DOS for the bilayer model
has a relatively complicated structure when $B=0$, with four distinct
energy scales (see \cite{Atkinson1999} for a complete discussion).
First, the coherence peaks associated with plane superconductivity are
split by chain-plane coupling (A).  A similar splitting has been
observed in STM experiments\cite{Ngai2007}.  We identify the
lower-energy of the two coherence peaks as the ``large gap''
$E_\mathrm{L}$. Because the chains act as pair-breakers, $E_\mathrm{L}$ is
slightly smaller than the gap in the single-layer model.  Next,
there are two gap energy scales associated with induced
superconductivity in the chains.  The larger of these (B) comes from
the region of the chain-derived Fermi surface nearest $(\pi,0)$ in the
Brillouin zone.  Here, the energetic proximity of the chain and plane
bands leads to a large hybridization of their wavefunctions and a
relatively large induced gap.  We believe that this energy scale,
denoted $E_\mathrm{s}$, is the small gap that has been seen in STM
experiments in zero field.  The smallest energy scale (C) is
associated with superconductivity on the portion of the chain Fermi
surface nearest the Brillouin zone centre.  This gap is wiped out by
small amounts of disorder or thermal broadening, and the proximity
model therefore predicts a finite residual density of states at the
Fermi energy, in accordance with tunneling
experiments.\cite{Valles1991,Yeh2001,Ngai2007}

The finite-field DOS is shown in Figure \ref{fig:cmpr}(b).  The field
washes out the distinct energy scales; however, $E_\mathrm{s}$ shows
up in the field-dependence of the DOS.  Compared with the single-layer
model, $\rho(\varepsilon_F)$ for the bilayer model is a strong
function of $B$ for $B<B^\ast$, where $B^\ast\sim
E_\mathrm{s}^2$\cite{Atkinson2008} is the crossover field mentioned
above.  For $B>B^\ast$, $\rho(\varepsilon_F)$ varies at a similar rate
with $B$ in both the single-layer and bilayer models.  Similar
behaviour has been predicted in an anisotropic single-band model for
\ybco\cite{Whelan2000}.  In the bilayer model studied here, $B^\ast$
is the magnetic field above which the vortex cores in the chains begin
to overlap.  From the perspective of the chains, there are therefore
distinct low- ($B<B^\ast$) and high- ($B>B^\ast$) field regimes.
Experimentally, $B^\ast \approx 1$ T in $\mathrm{YBa_2Cu_3O_{6.95}}$.
Because we have chosen a large value for the chain-plane
coupling,
the data shown here for the
$60\times 60\times 2$ lattices belong to the regime $B<B^\ast$.

\begin{figure}
\begin{center}
\includegraphics[width=0.6\columnwidth]{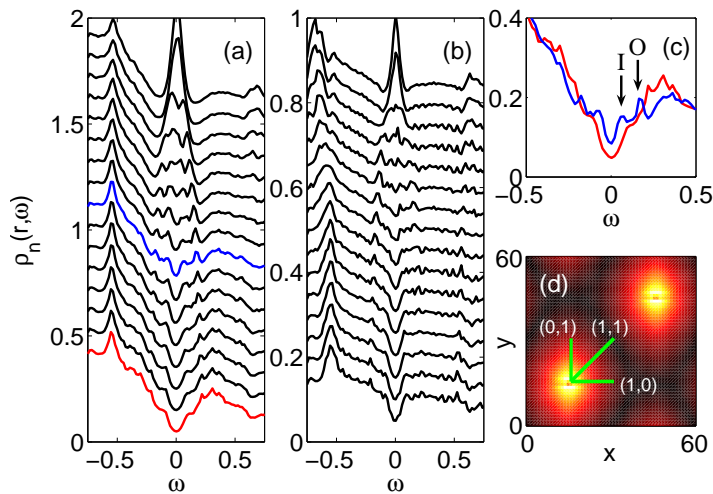}
\caption{Local density of states in the bilayer model. (a) The LDOS
$\rho_1(\rr,\omega)$ in the 2D plane at positions between the vortex
core (top curve) and the edge of the vortex (bottom curve).  Curves
are offset for clarity.  Positions are taken along the $(1,0)$ cut
shown in (d).  (b) $\rho_2(\rr,\omega)$ in the 1D chain layer at
identical positions to (a).  (c) Two of the LDOS curves from (a) are
compared, showing the dispersion of the inner (I) and outer (O) peaks.
Curve colours are the same as in (a).  (d) Spatial dependence of the
current density $|{\bf j}_1(\rr)|$ in the plane layer for a pair of
vortices.  The lines indicate the cuts along which LDOS curves are
displayed in this figure [$(1,0)$], Figure \protect\ref{fig:cut3}
[$(1,1)$] and Figure \protect\ref{fig:cut4} [$(0,1)$].  Note that
the $(0,1)$ and $(1,0)$ cuts are parallel and perpendicular to the chains
respectively.}
\label{fig:cut2}
\end{center}
\end{figure}

\begin{figure}
\begin{center}
\includegraphics[width=0.6\columnwidth]{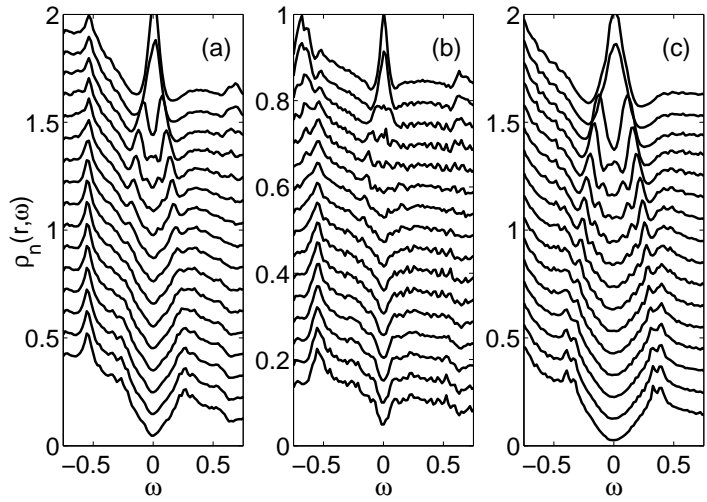}
\caption{Local density of states in the bilayer model for the (a)
plane and (b) chain layers.  Cuts are taken along the $(1,1)$
direction, illustrated in Figure \protect\ref{fig:cut2}(d), with the
vortex core corresponding to the top curve.  The LDOS is also shown in
(c) for the single-layer model.}
\label{fig:cut3}
\end{center}
\end{figure}

\begin{figure}
\begin{center}
\includegraphics[width=0.6\columnwidth]{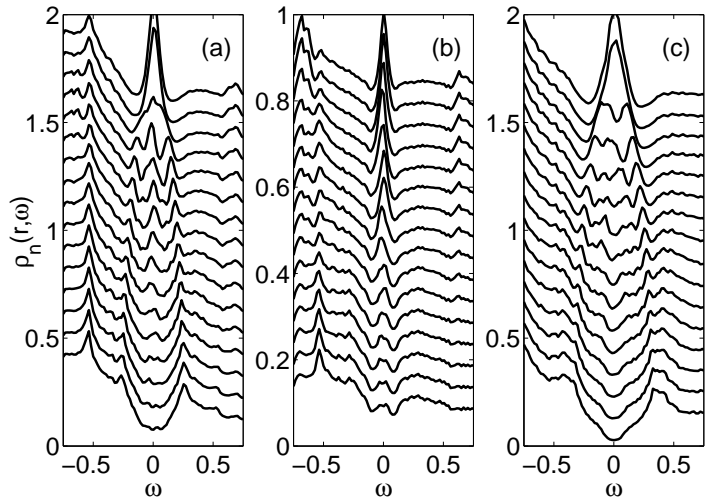}
\caption{Local density of states in the bilayer model for the (a)
plane and (b) chain layers.  Cuts are taken along the $(0,1)$
direction, illustrated in Figure \protect\ref{fig:cut2}(d), with the
vortex core corresponding to the top curve.  The LDOS is also shown in
(c) for the single-layer model.}
\label{fig:cut4}
\end{center}
\end{figure}

The LDOS for the bilayer is shown along a series of cuts through the
vortex along the $(1,0)$ direction in Figure \ref{fig:cut2}, the
$(1,1)$ direction in Figure \ref{fig:cut3}, and the $(0,1)$ direction
in Figure \ref{fig:cut4}.  The cuts are illustrated in Figure
\ref{fig:cut2}(d).  There are three sources of anisotropy which
contribute to differences in $\rho_n(\rr,\omega)$ along the three
directions.  First, the $d$-wave order parameter vanishes for
quasiparticles travelling in the $(1,\pm 1)$ directions and obtains
its maximum along the $(1,0)$ and $(0,1)$ directions. Second, the
shortest path between vortices lies along the $(1,\pm 1)$ directions.
Third, the chains run parallel to the $(0,1)$ direction.  We can
separate the third factor from the first two by comparing our bilayer results
with those for the single-layer model.

First, we consider the cut along the $(1,0)$ direction (Figure
\ref{fig:cut2}).  The LDOS in the plane layer exhibits one notable
difference from the single layer case shown in Figure \ref{fig:cut1}:
the ZBCP splits into a quartet, instead of a pair, of dispersing peaks
as one moves away from the vortex core.  The outer peaks asymptotically
approach $E_\mathrm{L}$ at large distances from the vortex core, while the
inner peaks asymptotically approach $E_\mathrm{s}$.  It appears as if
the presence of two superconducting energy scales leads to the formation
of two distinct sets of dispersing quasi-bound resonances.

The LDOS curves shown in Figure \ref{fig:cut2}(b) come from different
chains at increasing distances from the vortex core.  As in the plane,
there is a ZBCP at the vortex core.  The splitting of the ZBCP that is
readily apparent in the plane layer is also present, although
difficult to resolve, in the chain layer.  The LDOS curves far from
the vortex core closely resemble the DOS for the chains when $B=0$:
there is a pronounced suppression of the DOS for
$|\omega|<E_\mathrm{s}$, a large residual DOS at
$\omega=\varepsilon_F$, and there are weak remnants of the coherence
peaks at $E_\mathrm{L}$ coming from the superconductivity in the
planes.

The LDOS is shown along the $(1,1)$ direction in Figure
\ref{fig:cut3}.  In both the single-layer and bilayer models, the
peaks disperse more quickly along the $(1,1)$ direction than along the
$(1,0)$ direction.  Furthermore, in the case of the bilayer model, the
inner peaks are weak and difficult to resolve.

When the cut is taken along the $(0,1)$ direction (Figure
\ref{fig:cut4}), the single layer model gives the same LDOS curves as
in the $(1,0)$ direction, as required by the fourfold symmetry of the
lattice.  The bilayer model, however, is qualitatively different along
the $(0,1)$ and $(1,0)$ cuts.  In the plane layer, there are a pair of
dispersing peaks that asymptotically approach $\pm E_\mathrm{L}$ at
large distances from the vortex core.  These peaks correspond to the
outer peaks found along the $(1,0)$ direction.  In contrast, the inner
peaks that are seen along the $(1,0)$ direction are absent here.
Instead, there is a pronounced ZBCP that extends $\sim 10$ unit cells
from the vortex core along the $(0,1)$ direction.  The ZBCP is
particularly pronounced in the chain layer, where extends a longer
distance than in the plane layer.  Closer examination reveals a very
slight splitting of the peak at large distances.  For reasons
discussed below, we believe that this splitting occurs because
chain-plane coupling makes the chain-derived states weakly 2D.

\begin{figure}
\begin{center}
\includegraphics[width=0.6\columnwidth]{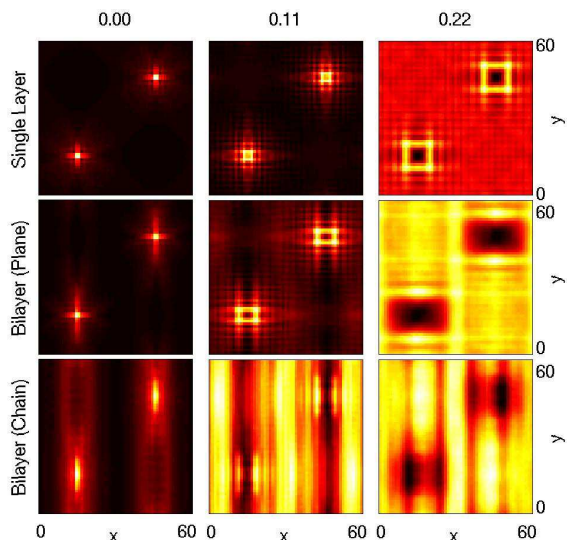}
\caption{LDOS maps for two vortices on a $60\times 60\times N_z$
lattice.  Figures show $\rho_n(\rr,\omega)$ as a function of $\rr$ for
the single-layer ($N_z=1$; top row), and bilayer model ($N_z=2$;
middle and bottom rows).  Each column corresponds to a different value of
$\omega$, indicated at the top of the column.}
\label{fig:map}
\end{center}
\end{figure}

For comparison, the spatial maps of $\rho_n(\rr,\omega)$ are shown at
three different values of $\omega$ in Figure \ref{fig:map}.  This
figure provides a different perspective on the results presented in
the preceding discussion.  We begin with the plots for
$\omega=\varepsilon_F$. The single-layer model exhibits a fourfold
symmetry, as expected.  The ZBCP appears as the bright cross in the
middle of each vortex core, and disappears away from core, where the
peaks disperse to higher energies.  In the bilayer model, we can
clearly see that the LDOS in the plane, $\rho_1(\rr,\varepsilon_F)$,
is orthorhomically distorted and has long tails along the chain
direction.  In the chain layer, the vortex core is even more
anisotropic; the extent of the core states along the $a$-axis is
similar to that in the plane layer, but the extent along the $b$-axis
is determined by the coherence length in the chains, $\xi_c = \hbar
v_\mathrm{F,c}/\pi E_s \approx 10 a_0$, where $v_\mathrm{F,c}$ is the
chain Fermi velocity.

At the two larger values of $\omega$, the LDOS shows a vortex core
that appears to expand with increasing $\omega$.  This is because the
apparent border of the core is determined by the positions of the
dispersing peaks at that value of $\omega$.  As discussed earlier, the
position and energy of these peaks are approximately linearly related.
The most noticeable difference between the single-layer and bilayer
models is the orthorhombic distortion of the vortex core in the
bilayer model, not present in the single-layer calculation. The
single-layer model exhibits the fourfold ``star'' shape, first
described in \cite{Schopohl1995}, while the LDOS in the bilayer plane
is stretched along the $a$-axis, perpendicular to the chains.  We note
that STM experiments in \ybco{} found a similar stretching of the
cores along the $a$-axis, with a ratio of $\sim 1.5$ between major and
minor axes.\cite{Fischer2007} Although our results are suggestive, it
is premature to declare the chains to be the underlying mechanism for
this anisotropy, especially given the number of unresolved
discrepancies between theory and experiment.  We return to this point
below.

We can construct a qualitative argument for the anisotropy between the
$(1,0)$ and $(0,1)$ directions in the dispersion of the inner peaks.
The fact that these peaks asymptotically approach $E_\mathrm{s}$ at
large distances from the vortex core suggests that they are closely
connected to superconductivity in the chain layer.  We therefore
consider the behaviour of quasiparticles travelling at the chain Fermi
velocity $v_\mathrm{F,c}$ within a chain a perpendicular distance
$r_\perp$ from the vortex core.  Since the direction along which the
quasiparticle travels does not change as it moves along the chain, the
quasiparticle angular momentum is conserved and is $L_\mathrm{c}=
r_\perp v_\mathrm{F,c}$.  This is obviously an approximate statement
since a quasiparticle, as a result of chain-plane hybridization, spends
a fraction of its time in the plane layer where its
trajectory is altered by the magnetic field.  The linear
relationship between $r_\perp$ and $L_\mathrm{c}$ is therefore also
approximate.  However, it is sufficient to explain the anisotropy
between $(1,0)$ and $(0,1)$ directions. We recall that in
conventional superconductors there is a linear relationship between
the angular momentum of a quasiparticle eigenstate and its
energy.\cite{Caroli1964,Gygi1991} For the $(1,0)$ cut, $r_\perp$ is
equal to the distance $r$ from the vortex core, $L_\mathrm{c}$ grows
linearly with $r$, and the inner peaks are expected (by this argument)
to disperse linearly with $r$.  This prediction is consistent with
the numerical results shown in Figure \ref{fig:cut2}.  On the
other hand, $r_\perp=0$ everywhere along the $(0,1)$ cut, since that
cut follows along the chain that passes through the vortex core.  We
therefore have $L_\mathrm{c}=0$ everywhere along this cut, and the
inner peaks remain at $\omega=\varepsilon_F$.  Since the
single-electron hopping between the chain and plane layers makes the
chain-derived states weakly 2D, $L_\mathrm{c}$ is only approximately
conserved along the length of the chain.  We therefore expect a very
weak dispersion of the peak energies along the $(0,1)$ direction.  
Again, this simple argument is consistent with numerical results,
shown in Figure \ref{fig:cut4}.

We finish this section with a brief discussion of the relevance of
this work to experiments.  There is a large body of literature
attempting to explain STM experiments in HTS, both in zero magnetic
field and in finite fields\cite{Fischer2007}. Existing theories are
successful at reproducing broad qualitative features of the DOS
spectrum, but generally fail to correctly predict quantitative details
such as the spatial dependence of the LDOS near isolated impurities or
vortex cores.  The same appears to be true of the work discussed here:
the proximity model is consistent with experiments indicating the
existence of a small superconducting gap scale $E_\mathrm{s}\approx 5$
meV but, as with previous calculations, there is a discrepancy between
the experiments and theory regarding the spatial dependence of the
LDOS.  Whether the failure to explain experiments is an experimental
issue, or is due to unknown tunneling matrix element effects, surface
states, or a failure of the basic mean-field model, is unclear at this
stage.

Because of these uncertainties, we seek simple qualitative predictions
stemming from the proximity model that might reveal something about
the electronic structure of the vortex cores.  We believe that one
qualitative feature predicted by the proximity model, the existence of
an anistropy in the low energy LDOS between the $(1,0)$ and $(0,1)$
directions, should hold regardless of the details of the model.  An
experiment to measure such an anisotropy would establish whether
$E_\mathrm{s}$ does, in fact, originate with the chains.  In order to
address this question, it will be necessary to perform experiments at
strong fields where the vortex lattice is square.

\section{Conclusions}
We have studied the local density of states near
a vortex core within a proximity model for \ybco.  This model
incorporates both CuO$_2$ planes and CuO chains.
The proximity model predicts that the CuO chains introduce
a set of dispersing peaks in the LDOS, in addition to the peaks
predicted for a single-layer superconductor.  These peaks are
associated with the small induced gap in the chain layer, and
consequently have a strongly anisotropic dispersion.  At
$\omega=\varepsilon_F$, the LDOS is extended along the chain
direction, but at $\omega>\varepsilon_F$, the vortex cores appear
stretched along the $a$-axis.

\ack
This work was supported by NSERC of Canada.  Calculations were
performed using the High Performance Computing Virtual Laboratory (HPCVL).

\section*{References}

\bibliographystyle{unsrt}
\bibliography{Atkinson_ybco}

\end{document}